\documentstyle[preprint,aps]{revtex}

\newtheorem{theorem}{Theorem}
\newtheorem{acknowledgement}[theorem]{Acknowledgement}

\begin{document}
\title{Thermal conductivity of MgB$_{2}$ in the superconducting state}
\author{M.Putti, V.Braccini, E.Galleani d'Agliano, F.Napoli, I.Pallecchi, A.S.Siri}
\address{INFM-LAMIA/CNR, Dipartimento di Fisica, Via Dodecaneso 33, 16146 Genova,\\
Italy}
\author{P.Manfrinetti, A.Palenzona,}
\address{INFM, Dipartimento di Chimica e Chimica Industriale, Via Dodecaneso 31,\\
16146 Genova, Italy}
\date{\today }
\maketitle

\begin{abstract}
We present thermal conductivity measurements on very pure and dense bulk
samples, as indicated by residual resistivity values as low as 0.5 mW cm and
thermal conductivity values higher than 200 W/mK. In the normal state we
found that the Wiedemann Franz law, in its generalized form, works well
suggesting that phonons do not contribute to the heat transport. The thermal
conductivity in the superconducting state has been analysed by using a
two-gap model. Thank to the large gap anisotropy we were able to evaluate
quantitatively intraband scattering relaxation times of $\pi $ and $\sigma $
bands, which depend on the disorder in different way; namely, as the
disorder increases, it reduces more effectively the relaxation times of $\pi 
$ than of $\sigma $ bands, as suggested by a recent calculation \cite{[1]}.
\end{abstract}

\bigskip {\bf 1.INTRODUCTION}

Since the discovery of $MgB_{2}$ \cite{[2]}, experimental findings
established a phonon mediated s-wave superconductivity. Despite its standard
origin, superconductivity in $MgB_{2}$ showed several unusual properties
that can be ascribed to the presence of two gaps with different amplitudes
emphasized by tunnelling and specific heat measurements. Theoretical studies 
\cite{[3]}\cite{[4]} pointed out that the peculiar electronic structure is
the origin of this behaviour, being the larger gap, $\Delta _{\sigma }$,
associated with two-dimensional $\sigma $ bands and the smaller one, $\Delta
_{\pi }$, associated with three-dimensional $\pi $ bands. The two-gap model
offers a simple explanation of several anomalies in the superconducting
state \cite{[5]}\cite{[6]}\cite{[7]}. Multiband effects due to the different
parity of $\sigma $ and $\pi $ bands were predicted in the normal state, too 
\cite{[1]}, and, recently, have found confirmation in transport measurements 
\cite{[8]}. Since interband impurity scattering turns out to be negligible,
different bands behave as separate conduction channels in parallel and
either $\sigma $ or $\pi $ channel prevails, depending on the disorder
degree, being the clean (dirty) samples dominated by $\pi $($\sigma $)
conduction.

In this paper we perform a quantitative analysis of thermal conductivity
measurements in the superconducting state to investigate the role of $\sigma 
$ and $\pi $ bands depending on the sample purity.

Thermal conductivity $\kappa $ in the superconducting state among the other
transport properties gives information on quasi-particle (QP) excitations
and their dynamics with the advantage of probing only the QP response, since
the superfluid does not carry heat. On the other hand, a major complication
in the analysis of the thermal conductivity is often a substantial phonon
contribution to the heat current. Consequently, the interpretation of
experimental data can be ambiguous.

From a basic point of view, QP condensation below $T_{C}$ causes a
decreasing of the electron contribution to the thermal conductivity, $\kappa
_{e}$, and an increasing of the phonon contribution, $\kappa _{p}$; thus $%
\kappa $ below $T_{C}$ can show a shoulder or a peak depending whether the
heat current is dominated by electrons or phonons, and a more complex
behaviour is exhibited when both the contribution are important.

From the beginning the thermal conductivity of $MgB_{2}$ showed the
unexpected feature of not exhibiting any signature of the superconducting
transition \cite{[9]}\cite{[10]}\cite{[11]}\cite{[12]}. This ''anomaly'' was
initially ascribed to a perfect compensation of $\kappa _{e}$ and $\kappa
_{p}$ \cite{[10]} or to the presence of large thermal resistance at the
grain boundaries \cite{[11]}. As a matter of fact, the discovery that $%
MgB_{2}$ presents two gaps, one of which $\Delta _{\pi }$ is so small that
QP condensation becomes exponential only below a reduced temperature of the
order of $t=T/T_{C}\sim 0.2$, opened new perspectives in the interpretation
of thermal conductivity data. Hence, the two gap model successfully used to
fit specific heat data \cite{[5]} can also be applied to the thermal
conductivity to estimate the fraction of energy carried by the two bands,
providing a useful mean to investigate multiband effects.

\bigskip

{\bf 2.EXPERIMENTAL DETAILS}

Dense (up to 2.4 g/cm$^{3}$, 90\% of the theoretical density), clean and
hard cylinder shaped samples have been prepared by a single step method\cite
{[13]} similar to the one reported in ref \cite{14-A}\cite{[14]}. Amorphous
or crystalline $B$ and $Mg$, put in $Ta$ crucibles welded under argon and
closed in quartz tubes under vacuum, were heated up to 950%
${{}^\circ}$%
C . An X rays spectrum of a sample prepared by crystalline $B$ is shown in
fig. 1. All the peaks associated with the $MgB_{2}$ phase are present, no
extra peaks due to the presence of free $Mg$ , $MgO$, are detected. In the
inset, SEM (Scanning Electron Microscopy) image of \ the same sample is
shown. The image shows a network of well connected grains (2-4 mm large).

Two specimens were prepared for physical measurements: one from crystalline
boron (MGB-1S) , one by using enriched $^{11}B$ (MGB11-1S). The samples were
cut in the shape of parallelepiped bar (1$\times $2-3$\times $12 mm$^{3}$).

The thermal conductivity was measured using a steady state flux method with
a heat flux sinusoidally modulated at low frequency ($\nu $=0.003-0.01 Hz ).
Under these conditions the thermal conductivity is extracted as $\kappa =$ $%
J(\nu )/\nabla T(\nu )$, where $J(\nu )$ is the heat flow provided at the
frequency $\nu $ and $\nabla T(\nu )$ is the temperature gradient
oscillating at the frequency $\nu $. A small resistive heater (1$\times $1 mm%
$^{2}$ ) is glued by GE varnish on the top end of the bar, being the bottom
of the bar thermally connected with the sample holder. In such way a
longitudinal heat flow is assured. The gradient applied to the sample was
varied from 0.1 to 0.3 K/cm. Seebeck effect was measured simultaneously with
the thermal conductivity providing a precise determination of the critical
temperature.

\bigskip

{\bf 3.RESULTS AND\ DISCUSSION}

Resistivity measurements of the samples from 30 to 300 K are shown in fig.
2. In table 1 we report $T_{C}^{onset}$, the amplitude of the transition $%
\Delta T_{C}$, $\rho (40K)$ and the residual resistivity ratio defined as $%
RRR=\rho (300K)/\rho (40K)$ for the two samples. The excellent quality of
the samples is proved by the high $T_{C}$ values (38.9, 38.7 K), the small $%
\Delta T_{C}$ values (0.2, 0.3 K), the low values of $\rho (40K)$ (0.6, 2.5 $%
\mu \Omega $cm) and the large values of $RRR$ (7-15). Between the two
samples the enriched $^{11}B$ (MGB11-1S) has the lower resistivity values
which can mainly be ascribed to the very good quality of the Eagle-Picher
enriched $^{11}B$ \cite{[14]}.

Thermal conductivity measurements of the two samples from 4 K to 250 K are
shown in fig. 3. The outstanding quality and high density of the samples is
evident from thermal transport properties as well. Indeed, these samples
have more than one order of magnitude higher thermal conductivities than
polycrystalline samples \cite{[9]}\cite{[10]}\cite{[11]}. In particular
MGB11-1S exhibits a thermal conductivity as large as 215 W/Km at 65 K which
is nearly two times higher than those of a single crystal \cite{[12]},
proving the excellent purity and density of this sample.

The thermal conductivity of MGB11-1S increases monotonically in the
superconducting state, a change of slope is observable at about 8 K, while
no signature of the superconducting transition is present; the normal state
curve exhibits a pronounced maximum at about 70 K. Similar behaviour is
presented by MGB-1S even if only data above 10 K are available.

The normal state behaviour of both samples is typical of a good metal in
which the electron contribution to heat transport prevails. In order to
estimate the relative weight of $\ \kappa _{e}$ and $\kappa _{p\text{ }}$we
can consider the effective Lorenz number defined as $L_{eff}=\kappa \rho /T$%
. This quantity is assumed equal to $L_{0}=2.45\times 10^{-8}W\Omega K^{-2}$
by the Wiedemann-Franz law (WFL); in good metals $L_{eff}$ is equal to $%
L_{0} $ at low temperature where the electron impurity scattering prevails,
then it decreases showing a minimum at about one tenth the Debye temperature 
$\Theta _{D}$, deeper and deeper with increasing sample purity \cite{[15]};
in dilute alloys where $\kappa _{p\text{ }}$ is not at all negligible $%
L_{eff}$ becomes larger than $L_{0}$ and the ratio $L_{eff}/L_{0}\simeq
(1+\kappa _{p}/\kappa _{e})$ gives information on the relative weight of $\
\kappa _{e}$ and $\kappa _{p\text{ }}$ \cite{[16]}. In fig. 4 we plot $%
L_{eff}$ for the two samples from 40 to 250 K. The curves exhibit the
typical behaviour of good metals with slightly different levels of purity:
at low temperatures the curves approach $L_{0}$ from below; they show a
minimum at around 130 K ($\sim 0.1\Theta _{D}$, considering that in $MgB_{2}$
$\Theta _{D}\sim $1000 K \cite{[17]}\cite{[18]}) which is more pronounced
for MGB11-1S; finally, the curves increase towards $L_{0}$. Thus, we can
conclude that in the normal state the WFL substantially works and $\kappa _{p%
\text{ }}$can be neglected. Similar results were obtained also in sintered
sample with low thermal conductivity \cite{[8]}, while different conclusions
have been drawn on a single crystal \cite{[12]}, where a violation of the
WFL was claimed. Actually, in a small sample with not well defined
geometrical shape as a single crystal the geometrical factor which relates
conductivity to conductance can be different in thermal and electrical
measurement, giving an uncorrect evaluation of the Lorenz number. In large
bar cut specimens the geometrical factors can be estimated with better
precision and more reliable verification of the WFL can be made.

In the superconducting state, since $\kappa _{p}$ decreases as $(T/\Theta
_{D})^{3}$, we can assume that the thermal conductivity is dominated by
electrons as well. Only at very low temperature $T\ll T_{C}$ due to QP
condensation which decreases $\kappa _{e}$ and enhances $\kappa _{p}$, their
relative weight can change. Hence, we assume that lattice vibrations give
negligible contribution to the thermal transport in the temperature region
of our interest and in the following we analyse the thermal conductivity
data in term of the electron contribution only.

The electron thermal conductivity in the superconducting state can be
written as:

\begin{equation}
\kappa _{e}^{s}(T)=\kappa _{e}^{n}(T)g(t,\sigma )
\end{equation}

where $\kappa _{e}^{n}(T)$ is the electron thermal conductivity in the
normal state and $g(t,\sigma )$, for a given reduced gap, $\sigma =\Delta
(0)/KT_{C}$, and reduced temperature, $t=T/T_{C}$ , takes into account the
QP condensation. The function $g(t,\sigma )$ was calculated in the framework
of the BCS theory in the dirty and clean limit cases \cite{[19]}\cite{[20]}
reproducing very well thermal conductivity in the low temperature
superconductors. To analyse experimental data using eq. (1) two main
problems have to be solved. First, eq. (1) has to be generalized to
multiband conduction; second, $\kappa _{e}^{n}(T)$ has to be estimated. We
start from the latter issue.

At low temperature, if the scattering with impurities prevails, $\kappa
_{e}^{n}(T)$ can be obtained by the WFL, $\kappa _{e}^{n}(T)=L_{0}T/\rho
_{0} $ where $\rho _{0}$ is the residual resitivity. But in clean samples
scattering with phonons has also to be considered. In ref. \cite{[8]} this
was done using a generalized WFL. We can write:

\begin{equation}
W_{e}=W_{e}^{i}+W_{e}^{p}
\end{equation}

where, for the Matthiessen's rule, the thermal resistance $W_{e}=1/\kappa
_{e}$ is the sum of the thermal resistivity for scattering with impurities, $%
W_{e}^{i}$, and for scattering with phonons,$W_{e}^{p}$. Following ref. \cite
{[15]} we can write:

\begin{equation}
W_{e}^{i}=\frac{\rho _{0}}{L_{0}T}
\end{equation}

\begin{equation}
W_{e}^{p}=\frac{\rho _{p}(T)\left[ \left( \frac{\Theta _{D}}{T}\right) ^{2}%
\frac{3}{\pi ^{2}}\left( \frac{n_{a}}{2}\right) ^{2/3}\right] }{L_{0}T}
\end{equation}

where $\rho _{p}(T)=4\rho ^{\prime }\Theta _{D}\left( \frac{T}{\Theta _{D}}%
\right) ^{5}\int_{0}^{\Theta _{D}/T}\frac{x^{5}dx}{(e^{x}-1)(1-e^{-x})}$ is
given by the Bloch-Gr\"{u}neisen equation, $\rho \prime $ is the temperature
coefficient and $n_{a}$ is the average number of electrons per unit cell in
a given band.

Despite its simplicity, this model describes very well the normal state
thermal conductivity of samples with very different degrees of disorder,
with reasonable parameter values \cite{[8]}; it contains the right features,
vanishing to zero at low temperature linearly with $T$ and requiring only
three free parameters ($\rho _{0}$, $\Theta _{D}$ and $C=4\rho ^{\prime
}\Theta _{D}\frac{3}{\pi ^{2}}\left( \frac{n_{a}}{2}\right) ^{2/3}$) whose
reliability can be checked by resistivity . The best fitting procedure is
performed from 40 K to 200 K and the curves obtained with the parameter
values listed in tab. 2 are reported in fig.4 as continuous lines. The
theoretical curves fit the experimental data in the normal state in
excellent way, while just below $T_{C}$, they lay above the data, decreasing
linearly with temperature, where the experimental data start to decrease
with larger slopes, due to the QP condensation. In tab. 2 we can see that
the $\Theta _{D}$ and $C$ values, which determine the intrinsic term , are
nearly the same for both the samples, while the $\rho _{0}$ values change by
a factor 4, as consequence of the different purity of the two samples;
moreover we can see that the $\rho _{0}$ values are slightly lower than $%
\rho (40K)$ reported in table 1. Anyway, $\rho _{0}$, $\Theta _{D}$ and $C$
values are in fair agreement with those obtained by fitting resistivity
measurements \cite{[8]}.

Now we address the problem of generalizing eq. (1) in the case of multiband
conduction. Following the approach used for the specific heat \cite{[5]}, we
can write:

\begin{equation}
\frac{\kappa _{e}^{s}(T)}{\kappa _{e}^{n}(T)}=xg(t,\sigma _{\pi
})+(1-x)g(t,\sigma _{\sigma })
\end{equation}

where $\sigma _{\pi }=\Delta _{\pi }(0)/KT_{C}$ and $\sigma _{\sigma
}=\Delta _{\sigma }(0)/KT_{C}$; the relative weights $x$ and $(1-x)$ that in
ref. \cite{[5]} are related to the energy which condenses in each band, in
our case rather represent the energy fractions carried by the $\pi $ and $%
\sigma $ bands, respectively; thus the relative weights take into account
the mobility of the carriers in each band, also.

If we indicate with $\kappa _{\pi }^{n}(T)=x\kappa _{e}^{n}(T)$ and $\kappa
_{\sigma }^{n}(T)=(1-x)\kappa _{e}^{n}(T)$ the thermal conductivity in the
normal state of $\pi $ and $\sigma $ bands, eq. (5) states that the $\pi $
and $\sigma $ bands conduction occurs in parallel, as suggested by ref. \cite
{[1]}.

Once a suitable g function is considered eq. (5) allows to fit the thermal
conductivity data in the superconducting state with $x,$ $\sigma _{\pi }$
and $\sigma _{\sigma }$ as free parameters. To describe $\sigma $ and $\pi $
QP we choose the $g$ function in the dirty limit \cite{[19]} in agreement
with the fact that below 40 K $\kappa _{e}^{n}(T)$ is dominated by the
scattering with impurities.

In figure 5 we show $\kappa /T$ in the superconducting state for the two
samples. The continuous line represents $\kappa _{e}^{n}(T)/T$, the dashed
line represents $\kappa _{e}^{s}(T)/T$ given by eq. (5). The parameters
obtained with the best fit procedure are summarized in table 2. We can see
that the two-gap model provides an excellent agreement with the experimental
data. We find $\sigma _{\pi }$=0.57 ( 0.6) and $\sigma _{\sigma }$=2.17
(1.9) for MGB11-1S (MGB-1S); these values, which correspond to $\Delta
(0)_{\pi }=1.9(2.0)$ $meV$ and $\Delta (0)_{\pi }=7.2(6.3)$ $meV$ are in
agreement with those found by specific heat data $\sigma _{\pi }\sim $%
0.6-0.65 and $\sigma _{\sigma }\sim $1.9-2.2. For the $x$ parameter we find
0.85 and 0.75 for MGB11-1S and MGB-1S, respectively. From the specific heat
analysis it comes out that the energy fraction which condenses in each band
is roughly the same; thus our results imply that carriers in $\pi $ bands
are more mobile than in $\sigma $ bands. The lacking signature of
superconducting transition in thermal conductivity, naturally follows from
the two-gap model; in fact, we find that the main contribution to heat
transport comes from carriers in $\pi $ bands whose strong condensation
starts for $\sigma _{\pi }/t>>1$ which means $t<0.2$. In MGB11-1S data a
change of slope occurs at about 8 K which in this framework represents the $%
\pi $ QP condensation; in fact in figure 5 the dotted line represents $%
\kappa _{e}^{s}(T)/T$ given by eq. (5) with $x=1$ and $\sigma _{\pi }$=0.57
. The curve fits well the data just below 10 K, while from 10 to 40 K also
the $\sigma $ QP excitations contribute to the transport.

The right weight of $\pi $ and $\sigma $ contributions to transport can be
calculated for both samples by $\kappa _{\pi }^{n}(T)=x\kappa _{e}^{n}(T)$
and $\kappa _{\sigma }^{n}(T)=(1-x)\kappa _{e}^{n}(T)$ once the reliability
of the $x$ and $\kappa _{e}^{n}(T)$ evaluation has been verified. We point
out that in the fit procedure, varying slightly the gap values, the quality
of the fit does not change very much, while the $x$ parameter for both the
samples is very well defined. On the other hand, below $T_{C}$, $\kappa
_{e}^{n}(T)=L_{0}T/\rho _{0}$ and therefore it only depends on the residual
resistivity $\rho _{0}$ which can be experimentally estimated. So we can
write:

\begin{equation}
\kappa _{\pi }^{n}(T)=L_{0}T\frac{x}{\rho _{0}}=L_{0}T\omega _{p_{\pi
}}^{2}\varepsilon _{0}\tau _{\pi }^{i}
\end{equation}

\begin{equation}
\kappa _{\sigma }^{n}(T)=L_{0}T\frac{1-x}{\rho _{0}}=L_{0}T\omega
_{p_{\sigma }}^{2}\varepsilon _{0}\tau _{\sigma }^{i}
\end{equation}

where $\varepsilon _{0}$ is the dielectric constant, $\omega _{p_{\pi
,\sigma }}$ are the plasma frequencies and $\tau _{\pi ,\sigma }^{i}$ are
the intraband scattering relaxation times with impurities for $\pi $ and $%
\sigma $ bands.

Our fits state that carriers in $\pi $ bands mainly contribute to carry
heat. Indeed, the plasma frequency is larger for $\pi $ than for $\sigma $
bands \cite{[6]} and this is enhanced in polycrystalline samples by
averaging in the three direction. On the other hand, the scattering
relaxation times change from sample to sample. Introducing in eq.s (6) and
(7) the values of $x$ and $\rho _{0}$ listed in tab. 2 and $\omega _{p_{\pi
}}$= 6.226 eV and $\omega _{p_{\sigma }}$=3.403 eV \cite{[6]} we obtain the
scattering relaxation times for the two samples summarized in tab. 3. For
MGB11-1S we find $\tau _{\pi }^{i}$=2.2$\times $10$^{-13}$ s and $\tau
_{\sigma }^{i}$=1.3$\times $10$^{-13}$ s and the residual mean free paths $%
l_{\pi ,\sigma }$=$\tau _{\pi ,\sigma }^{i}v_{F\pi ,\sigma }$ , where $%
v_{F\pi ,\sigma }$ are the averaged Fermi velocities for $\pi $ an $\sigma $
bands ($v_{F\pi }$=5.6$\times $10$^{5}$ m/s and $v_{F\sigma }$=3.2$\times $10%
$^{5}$ m/s \cite{[6]}), come out $l_{\pi }$=1.2 $\times $10$^{-7}$m and $%
l_{\sigma }$=4.1 $\times $10$^{-8}$ m. These values are very large,
exceeding the lattice constants by more than two orders of magnitude,
indicating the excellent purity of this sample. For MGB-1S the relaxation
rates are reduced, but the residual mean free paths are still large ($l_{\pi
}$=2.9 $\times $10$^{-8}$ m and $l_{\sigma }$=1.8 $\times $10$^{-8}$ m).

Now we can look at the relaxation times in more details. In MGB11-1S $\tau
_{\pi }^{i}$ 
\mbox{$>$}%
$\tau _{\sigma }^{i}$, while in MGB-1S both the relaxation times are lower
and $\tau _{\pi }^{i}$ 
\mbox{$<$}%
$\tau _{\sigma }^{i}$ . Practically, as the disorder increases, it reduces
more effectively $\tau _{\pi }^{i}$ than $\tau _{\sigma }^{i}$. Actually, it
is not easy to say which kind of disorder is present in MGB-1S, being itself
a quite pure sample. But typical defects in $MgB_{2}$ are vacancies or
substitutions in the $Mg$ site, which form more easily than in the $B$ site.
In these cases the relaxation rate for intraband impurity scattering is
larger in $\pi $ than in $\sigma $ bands ($\tau _{\pi }^{i}$ 
\mbox{$<$}%
$\tau _{\sigma }^{i}$) \cite{[1]}. This prediction, which was well verified
for large amount of defects \cite{[8]}, is here confirmed also going from
very pure to rather pure samples.

This result can open new perspectives in thermal transport properties. In
fact, as the disorder increases, we expect that the transport of $\sigma $
bands prevails, and in superconducting state, the rapid QP condensation due
to the large gap related to $\sigma $ bands can become evident. Thus, in
disordered samples, we expect that the thermal conductivity would diminish
in absolute values, but it should show a wide shoulder below $T_{C}$. This
fact, if true, is quite unusual, in fact the disorder generally smoothens,
rather than enhancing features.

As a matter of fact, looking at the thermal conductivity data in literature,
polycrystalline samples do not show more evident shoulder than clean
samples. We think that polycrystalline samples, that surely are more
disordered than single crystals or bulk samples, present resistive grain
boundaries, which contribute to the thermal resistance masking the intrinsic
behaviour of superconducting grains \cite{[11]}. To emphasize the transport
of carriers in $\sigma $ bands by the progressive inhibition of transport in 
$\pi $ bands, it is necessary to gradually introduce defects in pure
samples. This can be done by suitable chemical substitution with the
advantage of introducing a controlled amount of defects in a chosen site. Up
to now, substituted samples have been obtained by sintering powders
previously synthesized from the pure elements and such samples are not
suitable for transport measurements. A second way is to introduce disorder
by irradiation, but in this case the problem is to obtain a uniform defect
distribution.

In conclusion, the role of disorder in $MgB_{2}$, which multiband effects
make so peculiar, has been studied in pure samples by thermal transport
measurements. We analysed in detail the thermal conductivity in the
superconducting state, and thank to the large gap anisotropy we were able to
evaluate quantitatively the intraband scattering relaxation times of $\pi $
and $\sigma $ bands.

\begin{acknowledgement}
This work work is partially supported by the Istituto Nazionale per la
Fisica della Materia through the PRA UMBRA
\end{acknowledgement}

\bigskip

{\bf Figure captions}

Figure 1. X-Rays pattern diffraction of an enriched $^{11}B$ bulk sample.
The inset shows a SEM image of the same sample. The length scale of the
picture is indicated in the bottom.

Figure 2. Resistivity measurements from 40 to 300 K.

Figure 3. Thermal conductivity measurements of the samples from 10 K to 250
K: the best fitting curves obtained with the parameter values listed in tab.
2 are reported as continuous lines.

Figure 4. The effective Lorenz number $L_{eff}/L_{0}=\kappa \rho /(TL_{0})$
from 40 to 250 K.

Figure 5. $\kappa /T$ in the superconducting state. The continuous line
represents the calculated $\kappa _{e}^{n}/T$; the dashed line represents $%
\kappa _{e}^{s}/T$ given by eq. (5) with the parameter values summarized in
tab. 2; the dotted line represents $\kappa _{e}^{s}/T$ given by eq. (5) with 
$x=1$ and $\sigma _{\pi }$\ given in tab.2.

\bigskip

\bigskip {\bf Table captions}

Table 1. Critical temperature $T_{C}^{onset}$, amplitude of the transition $%
\Delta T_{C},$\ $\rho (40K)$ and residual resistivity ratio defined as $%
RRR=\rho (300K)/\rho (40K)$.

Table 2. Values of the parameters $\theta $, $\rho _{0}$ and $C$ obtained
from the normal state thermal conductivity best fit and $\sigma _{\pi }$, $%
\sigma _{\sigma }$ and $x$ obtained by fitting eq. (5) with the
superconducting state thermal conductivity.

Table 3. The intraband scattering time with impurities for $\pi $ and $%
\sigma $ bands, $\tau _{\pi }^{i}$ and $\tau _{\sigma }^{i}$and the residual
mean free paths $l_{\pi ,\sigma }$=$\tau _{\pi ,\sigma }^{i}v_{F\pi ,\sigma
} $ ($v_{F\pi }$=5.6$\times $10$^{5}$ m/s and $v_{F\sigma }$=3.2$\times $10$%
^{5}$ m/s are the averaged Fermi velocities for $\pi $ an $\sigma $ bands 
\cite{[6]}).

\bigskip

\bigskip

\ {\bf Table 1}

. 
\begin{tabular}{||l|l|l|l|l||}
\hline\hline
$sample$ & $T_{C}^{onset},K$ & $\Delta T_{C},K$ & $\rho (40K),\mu \Omega cm$
& $RRR$ \\ \hline
MGB11-1S & 38.7 & 0.2 & 0.58 & 15.3 \\ \hline
MGB-1S & 38.9 & 0.3 & 2.1 & 7.1 \\ \hline\hline
\end{tabular}

\bigskip

\bigskip

{\bf Table 2}

\bigskip 
\begin{tabular}{||l|l|l|l|l|l|l||}
\hline\hline
$sample$ & $\theta ,K$ & $\rho _{0},\mu \Omega cm$ & $C,\mu \Omega cmK^{-2}$
& $\sigma _{\pi }$ & $\sigma _{\sigma }$ & $x$ \\ \hline
MGB11-1S & 1190 & 0.50 & 12 & 0.57 & 2.17 & 0.85 \\ \hline
MGB-1S & 1130 & 1.9 & 18 & 0.60 & 1.90 & 0.75 \\ \hline\hline
\end{tabular}

\bigskip

{\bf Table 3}

. 
\begin{tabular}{||l|l|l|l|l||}
\hline\hline
$sample$ & $\tau _{\pi }^{i},s$ & $\tau _{\sigma }^{i},s$ & $l_{\pi },m$ & $%
l_{\sigma },m$ \\ \hline
MGB11-1S & 2.2$\times $10$^{-13}$ & 1.3$\times $10$^{-13}$ & 1.2$\times $10$%
^{-7}$ & 4.1$\times $10$^{-8}$ \\ \hline
MGB-1S & 0.5$\times $10$^{-13}$ & 0.6$\times $10$^{-13}$ & 2.9$\times $10$%
^{-8}$ & 1.8$\times $10$^{-8}$ \\ \hline\hline
\end{tabular}

\bigskip

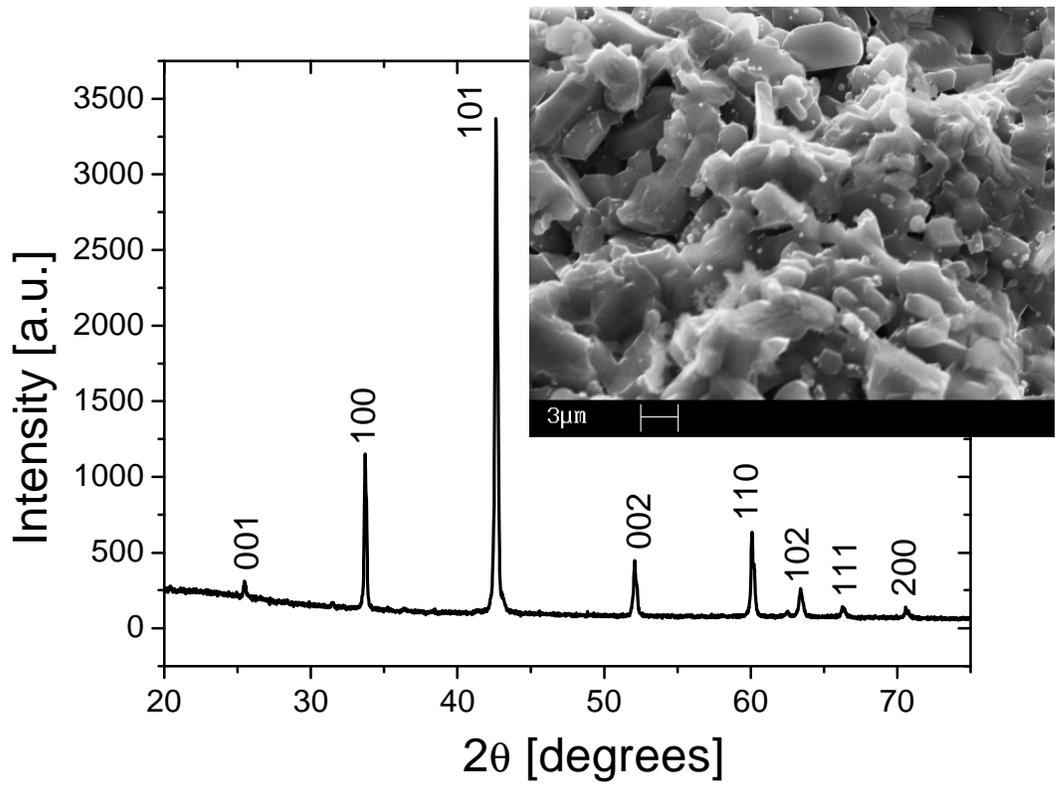
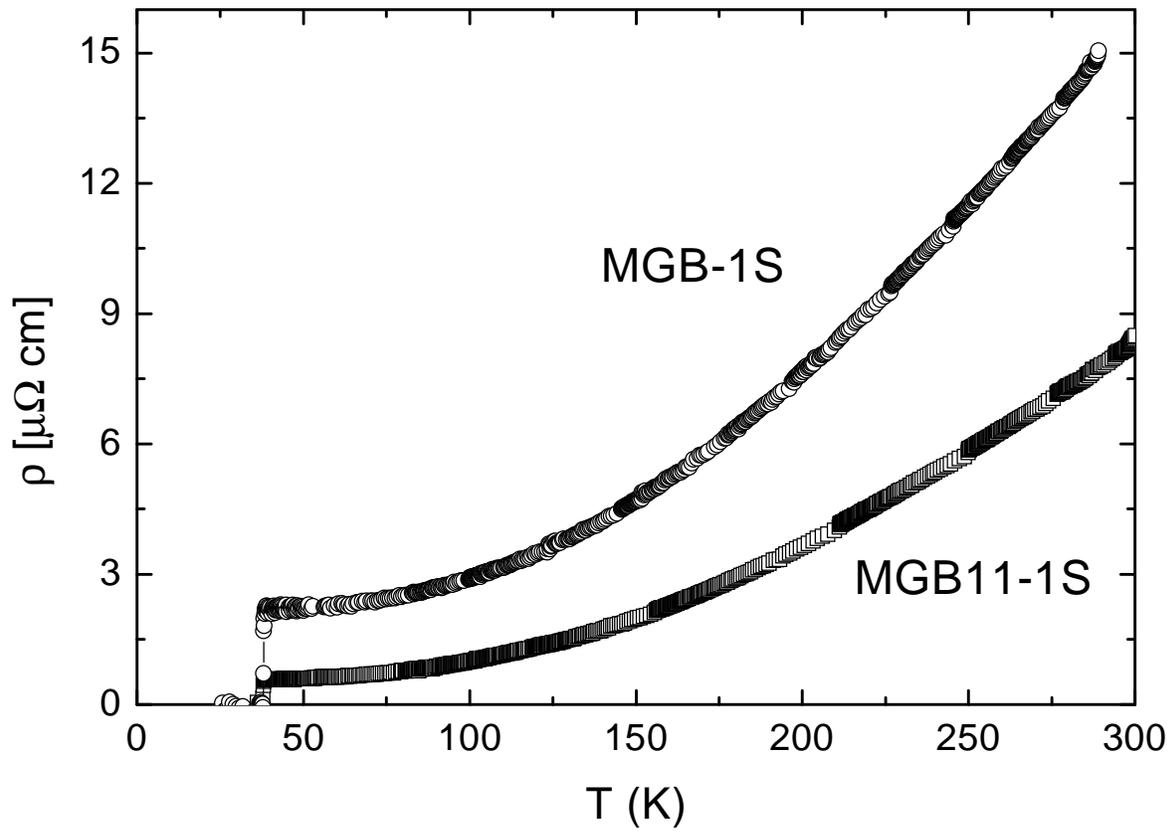
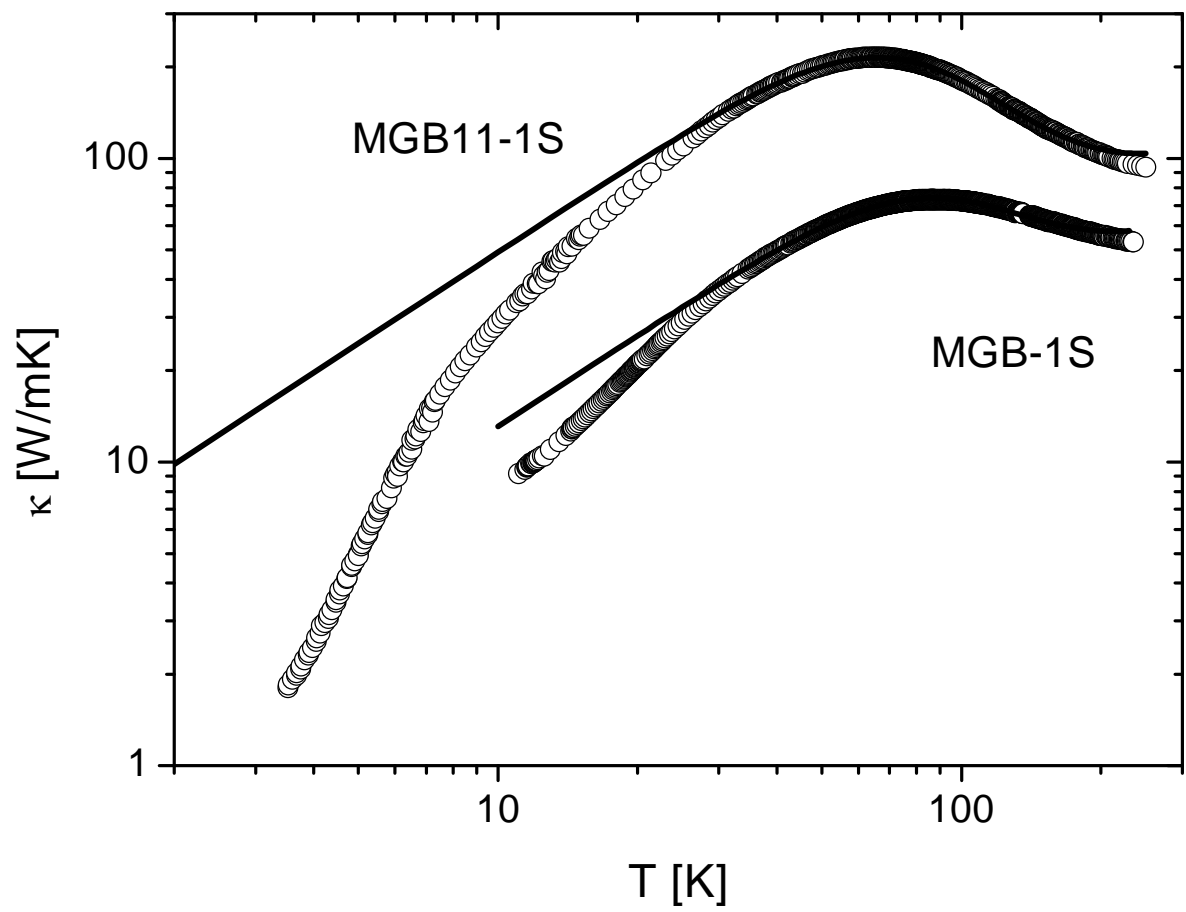
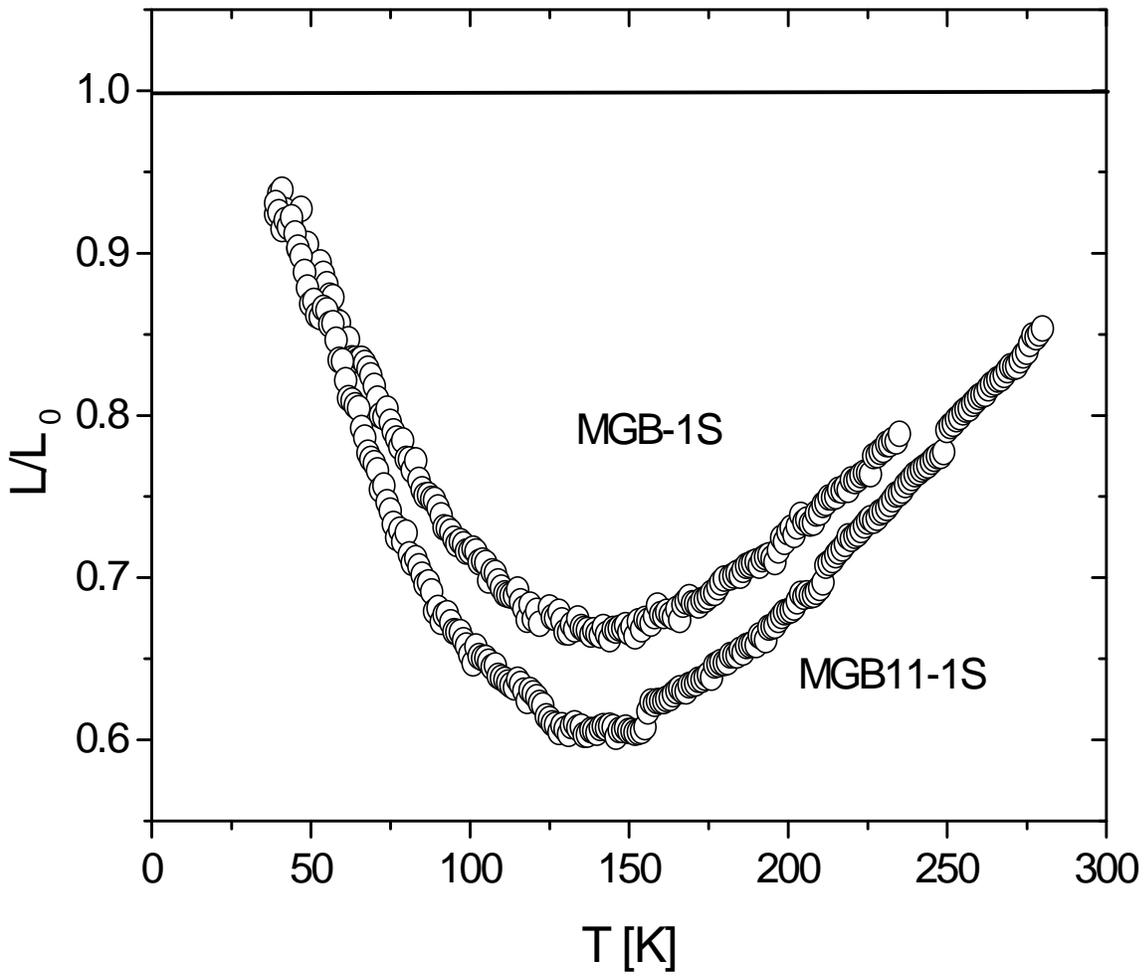
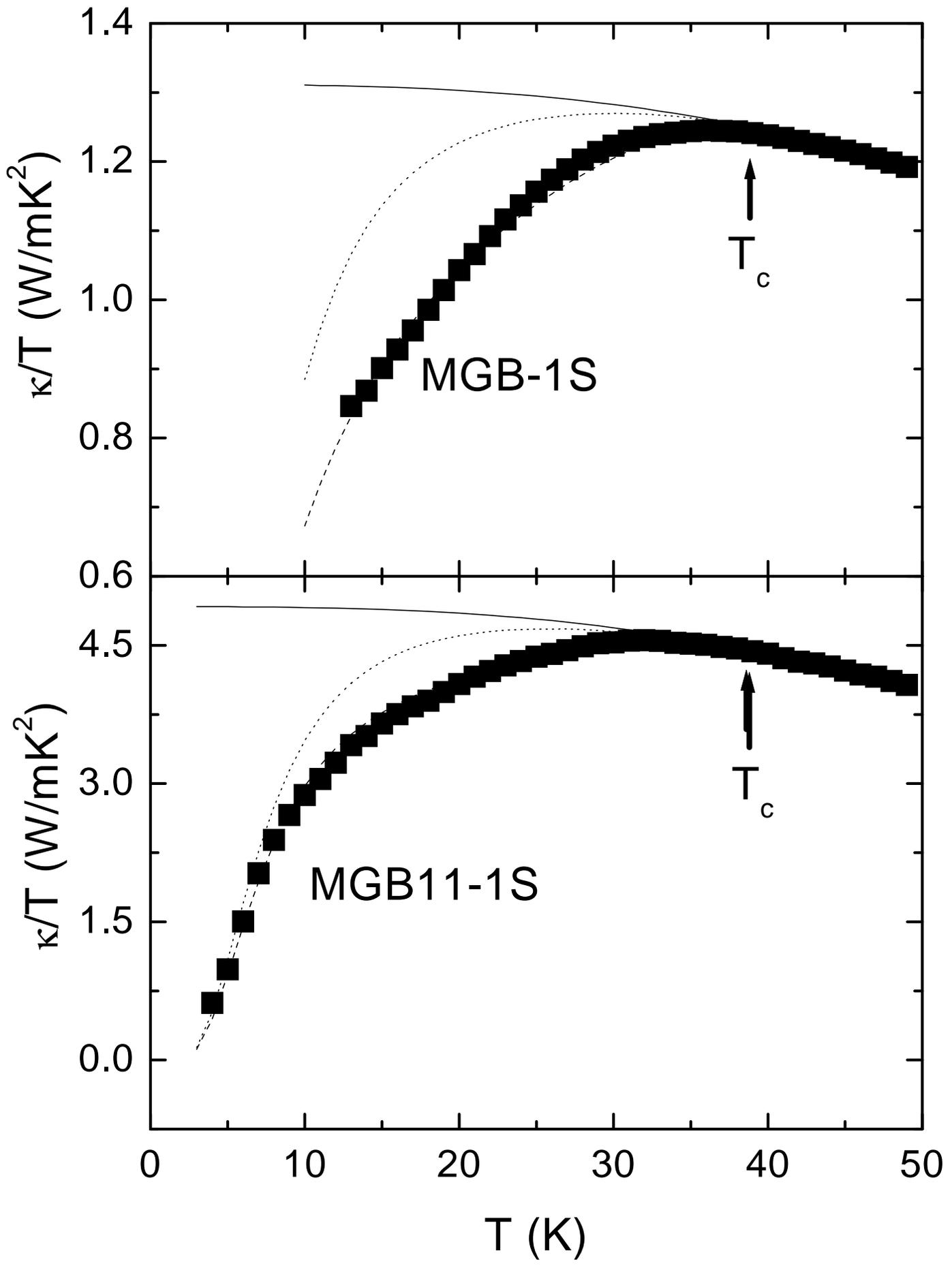
\end{document}